\documentclass[11pt,twoside]{article}

\usepackage{asp2006}
\usepackage{epsf}
\usepackage{graphicx}

\begin{document}
\title{Distinguishing Active Galactic Nuclei and Star Formation}   
\author{Brent Groves\altaffilmark{1} and Lisa Kewley\altaffilmark{2}}   
\affil{\altaffilmark{1}Sterrewacht Leiden, Leiden University, P. O. Box
9513, 2300 RA Leiden, The Netherlands}    
\affil{\altaffilmark{2}Institute for Astronomy, University of Hawaii, 2680 Woodlawn Drive, 
Honolulu, HI 96822, USA}

\begin{abstract} 
Using the large emission line galaxy sample from the Sloan Digital Sky
Survey we show that Star forming galaxies,
Seyferts, and low-ionization nuclear emission-line regions (LINERs)
form clearly separated branches on the standard optical diagnostic
diagrams. We derive a new empirical classification scheme which
cleanly separates these emission-line galaxies, using
strong optical emission lines. 
Using this classification we identify a few distinguishing host galaxy
properties of each class, which, along with the emission line
analysis, suggest continuous evolution from one class to another.
As a final note, we introduce models of both Starforming galaxies and
AGN narrow line regions which can explain the distribution of galaxies
on standard emission line ratio diagrams, and possibly suggest new
diagnostics across the emission spectrum.
\end{abstract}


\section{Introduction}

Since emission line objects were first observed, ratios of strong lines
have been to distinguish the various classes of emission line
nebulae and ionization mechanisms. One of the first line ratio diagrams, and
still one of the most powerful was that of 
[N{\sc ii}]$\lambda 658.4$/H$\alpha$ vs [O{\sc iii}]$\lambda
500.7$/H$\beta$, proposed by \citet{BPT81}, and commonly called the
BPT diagram. This diagram uses strong, optical lines of close
proximity in the ratios,  
limiting reddening and spectrophotometric effects, and is able to
clearly distinguish different classes of ionization. This and other
diagnostic diagrams along these lines were explored in detail later by
\citet[][(V\&O)]{Veilleux87}. However, with the advent of the Sloan Digital
Sky Survey (SDSS) we have been able fill this diagram and explore
details not seen before, enabling us to better understand the distribution and
extent of the emission line galaxies on this diagram.
 
\section{SDSS Emission Line Galaxies}

Within this work we used the SDSS Data Release 4 (DR4) spectroscopic galaxy
sample \citep{DR4}, which includes spectra of over 500,000 objects. 
The spectra are taken using
3-arcsec diameter fibers, positioned as close as possible to the
centers of the target galaxies. The flux- and wavelength-calibrated
spectra cover the range from 3800 to 9200\AA, with a resolution of
$R \sim$1800. 

At the median redshift of the sample ($z\sim 0.1$) the spectroscopic
fiber typically contains 20 to 40 percent of the total galaxy light,
thus contain a component due to the host galaxy as well
as any nuclear AGN source. 
As described in \citet{Tremonti04}, we subtract 
the contribution of the stellar continuum from each spectrum, using the
best fitting 
combination of template spectra from the population synthesis code of
\citet{Bruzual03}. The best-fitting model also can be used to estimate
stellar masses and star-formation histories \citep{Kauffmann03a}.

The sample criteria we use for the emission line galaxies is described
in detail in \citet{Kewley06}.  
The sample is limited to narrow emission line galaxies with 
redshifts above 0.02.               
We also apply a signal-to-noise cut ($S/N > 3$) on the
six dominant emission lines; 
H$\beta$ $\lambda 486.1$nm, [O{\sc iii}] $\lambda 500.7$nm, 
[O{\sc i}] $\lambda 630.0$nm, H$\alpha$ $\lambda 656.3$nm, 
[N{\sc ii}] $\lambda 658.4$nm, and the doublet [S{\sc ii}] $\lambda
671.6,3.1$nm.   
This gives a total sample of $\sim 85,000$ emission line galaxies.

\section{Emission Line Galaxy Classification}

The division of AGN and Starforming (SF) galaxies on the BPT diagram has
been known for a while, and the work of \citet{Kewley02} provided a
theoretical basis for the
distribution of starforming galaxies on this and the (V\&O) diagrams.
This theoretical work provided a ``maximal starburst'' line and clear
divisions between AGN and SF galaxies;
\begin{eqnarray}\label{eqn:niiK}
 \log([\mathrm{OIII}]/\mathrm{H}\beta) &=&
0.61/(\mathrm{NII}]/\mathrm{H}\alpha-0.47)) + 1.19,\\
\log([\mathrm{OIII}]/\mathrm{H}\beta) &=&
0.72/(\mathrm{SII}]/\mathrm{H}\alpha-0.32)) + 1.30,\\
\log([\mathrm{OIII}]/\mathrm{H}\beta) &=&
0.73/(\mathrm{OI}]/\mathrm{H}\alpha-0.59)) + 1.33,
\end{eqnarray}
with all AGN galaxies lying above this line.

After this work, a clearer delineation between AGN and SF galaxies on
the BPT diagram was 
seen by \citet{Kauffmann03b} when the large sample of SDSS galaxies
were plotted. A dividing line was fitted to these,  giving the
empirical relation, 
\begin{equation}\label{eqn:nii}
\log([\mathrm{OIII}]/\mathrm{H}\beta) =
0.61/(\mathrm{NII}]/\mathrm{H}\alpha-0.05)) + 1.3.
\end{equation}

With these diagnostics,the AGN branch, composed of Seyfert
galaxies, LINER-type galaxies and composite AGN/SF galaxies, still
remained mixed. However, when the V\&O [S{\sc ii}]/H$\alpha$ and
[O{\sc i}]/H$\alpha$ vs [O{\sc iii}]/H$\beta$ diagrams are plotted
with SDSS galaxies a clear minimum is seen in the AGN branch
\citep[see figure 3, ][]{Kewley06}. When
this minimum is fitted with a line in both diagrams, an empirical
division for Seyferts and LINERs is obtained,
\begin{eqnarray}\label{eqn:LINERs}
\log([\mathrm{OIII}]/\mathrm{H}\beta) &=&
1.89\log([\mathrm{SII}]/\mathrm{H}\alpha) + 0.76,\\
\log([\mathrm{OIII}]/\mathrm{H}\beta) &=&
1.18\log([\mathrm{OI}]/\mathrm{H}\alpha) + 1.30.
\end{eqnarray}

\begin{figure}
\center
\includegraphics[width=\hsize]{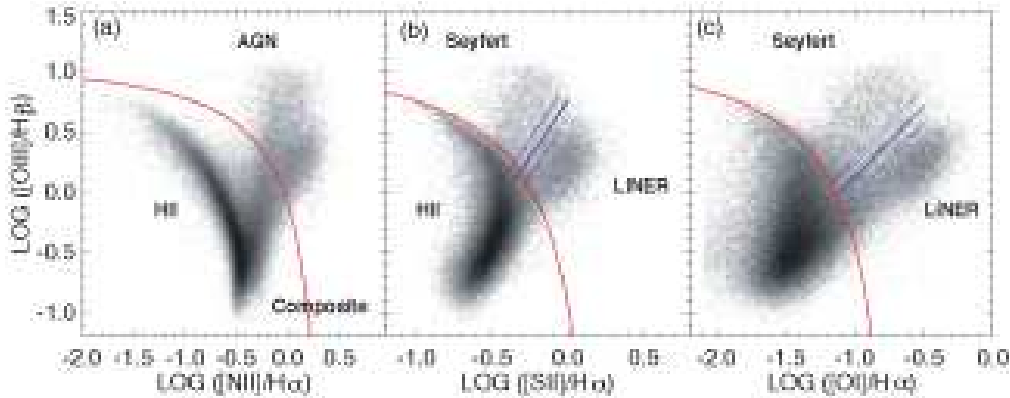}
\caption{The three emission line ratio diagnostic diagrams showing the 6
dividing lines. The solid curves in all three show the Kewley extreme
starburst definition, the dashed curve in (a) shows the Kauffmann
empirical relation, and the straight lines in (b) and (c) show the
LINER/AGN divisions.\label{fig:division}} 
\end{figure}

Together, these equations define 5 different classes of emission line
galaxies, as shown in figure \ref{fig:division}.

  Starforming galaxies, lying below the Kewley and Kauffmann curves,
      are the most numerous, making $\sim 64000$ (75\%) of the SDSS
      sample.

 Composite AGN/SF galaxies, lying between the Kauffmann and
      Kewley curves on the BPT diagram, are only 5900 galaxies, 7\% of
      the sample.

Seyfert galaxies, lying above the Kewley curves and the [O{\sc
      i}] and [S{\sc ii}] LINER lines, are only 3\% of the SDSS sample
      (2400).

LINERs, lying on the other side of the LINER lines to Seyferts
      are more numerous at 6000 galaxies (7\%)

 The remaining galaxies (8\%) fall in different classifications
      on at least one of the diagrams and are classified as ambiguous.

These classifications allow the differences and similarities between
Starforming galaxies and AGN to be considered.
This was first done by
\citet{Kauffmann03b}, and found that SF galaxies are
predominantly found in younger, more metal poor, later-type hosts than
AGN. However, as they cover a much larger range in metallicity than
AGN, their line properties are much more varied \citep{Groves06}.

This was extended in detail in the work reported here \citep{Kewley06}. 
The Composite galaxies, as would be expected for an intermediary
population, lie somewhere between AGN and metal-rich Starforming
galaxies in their host properties.
Seyferts tend to lie in the ``green valley'' in terms of their host
properties, having intermediary properties between the blue, SF
disc galaxies and the ``red \& dead'' ellipticals. 
LINERs
are older, more massive, less dusty, less concentrated, and they have
higher velocity dispersions and lower [O{\sc iii}] luminosities than Seyfert
galaxies have. Seyferts and LINERs are most strongly distinguished by
their [O{\sc iii}] luminosities. 
One remarkable property of the AGN is that at fixed L[O{\sc
iii}]/$\sigma_{*}^4$, 
which is an indicator of the black hole accretion rate relative to the
Eddington rate, all
differences between Seyfert and LINER host properties
disappear. LINERs and Seyferts form a continuous sequence, with LINERs
dominant at low L/L$_{EDD}$ and Seyferts dominant at high L/L$_{EDD}$. These
results suggest that the majority of LINERs are AGN and that the
Seyfert/LINER dichotomy is analogous to the high/low-state models and
show that pure LINERs require a harder ionizing radiation field with
lower ionization parameter than required by Seyfert galaxies,
consistent with the low and high X-ray binary states.

\section{A Model Understanding of the AGN \& SF Emission}

In addition to using the emission line diagrams to define
\emph{empirical} relationships and classifications, we can also use
these diagrams to better define our \emph{theoretical} models, which in turn
help our understanding of emission line objects. With the starburst models of
\citet{Dopita06}, we can clearly see that the dominating horizontal
extent of the SF galaxies in the BPT diagram is driven by metallicity
variations, while the vertical spread driven by a mix of stellar age
and ISM pressure.

Using the models of \citet{Groves04} we can see that the variation in
the AGN branch is driven by a combination of effects, 
predominantly by ionization and metallicity in the horizontal [N{\sc ii}]
direction, and ionization and SF/AGN mixing in the vertical.

Together with the observations these models can help us correctly
diagnose where exactly the ionizing photons are arising from in these
galaxies, along with some important host galaxy parameters.

\section{Conclusions}

The SDSS has given us a large sample of $z\sim 0.07$ emission line
galaxies, truly filling out the BPT line ratio diagram. This has
enabled clear, empirical classifications of Emission-Line galaxies to made, and
an exploration of host properties of these different classes. These
classifications and properties suggest some form of evolution, from the actively
starforming galaxies across the AGN Seyfert phase to the
final, elliptical LINER phase.
Along with new self-consistent theoretical models, the spread in this and other diagnostic
diagrams is now being understood, allowing a full classification based
on both observation and theoretical knowledge.



\end{document}